\documentclass[prd,preprint,
superscriptaddress,showpacs,nofootinbib,%
tightenlines,
]{revtex4}
\usepackage{epsfig}
\usepackage{amssymb}

\newcommand{\ben}{\begin{displaymath}}
\newcommand{\een}{\end{displaymath}}
\newcommand{\be}{\begin{equation}}
\newcommand{\ee}{\end{equation}}
\newcommand{\bea}{\begin{eqnarray}}
\newcommand{\eea}{\end{eqnarray}}
\begin{document}
\title{Ostrogradsky's Hamilton formalism and quantum corrections}
\author{J.~Gegelia}
\affiliation{Institut f\"ur Kernphysik, Johannes
Gutenberg-Universit\"at, D-55099 Mainz, Germany} \affiliation{High
Energy Physics Institute of TSU, 0186 Tbilisi, Georgia}
\author{S.~Scherer}
\affiliation{Institut f\"ur Kernphysik, Johannes
Gutenberg-Universit\"at, D-55099 Mainz, Germany}
\date{March 23, 2010}
\begin{abstract}

   By means of a simple scalar field theory it is demonstrated that
the Lagrange formalism and Ostrogradsky's Hamilton formalism
in the presence of higher derivatives, in general, do not lead to
the same results.
   While the two approaches are equivalent at the classical level,
differences appear due to the quantum corrections.
\end{abstract}


\pacs{
04.60.Ds,
04.60.Gw,
03.70.+k}



\maketitle

\section{Introduction}

   The higher-derivative dynamics is of particular interest
in the context of modern effective quantum field theories
(see, e.g., Refs.~\cite{Morozov:2007rp,Wise:2009mi} and
references therein).
   However, the quantization of Lagrangians involving
higher derivatives is a non-trivial problem.
   The canonical quantization of field theories using the Hamilton
formalism is a reliable method leading to a unitary scattering matrix.
   At the classical level, the Hamilton formalism for Lagrangians with
higher derivatives was developed by Ostrogradsky \cite{Ostrogradsky:1850}
a long time ago.
   The canonical quantization based on Ostrogradsky's Hamilton formalism
can be found, e.g., in Ref.~\cite{gitman}.

   In this work we examine Ostrogradsky's Hamilton formalism and
demonstrate that this method, although equivalent to the Lagrange
formalism at the classical level, may lead to wrong results due
to the quantum corrections.

\section{A Toy Model}
\label{toymodel}
   We consider the following Lagrangian of two scalar fields $A$ and $\Phi$,
\begin{equation}
{\cal L}_1(A,\Phi)=\frac{1}{2}\partial_\mu A\partial^\mu A
+ \frac{1}{2}\partial_\mu\Phi\partial^\mu\Phi-\frac{M^2}{2}\,\Phi^2\,,
\label{toylagr1}
\end{equation}
describing free massless ($A$) and massive ($\Phi$) spinless particles.
   For simplicity, we do not include the dependence on the partial
derivatives of the fields in the list of arguments of the Lagrangian.
   The momenta canonically conjugated to the fields $A$ and $\Phi$ are
defined by
\begin{eqnarray}
\label{pA1}
p_A&=&\frac{\partial {\cal L}_1}{\partial\, \partial_0 A}=\partial_0 A\,,\\
\label{pPhi1}
p_\Phi&=&\frac{\partial {\cal L}_1}{\partial\, \partial_0 \Phi}=\partial_0 \Phi\,,
\end{eqnarray}
resulting in the Hamiltonian
\begin{equation}
\label{hamiltonianh1}
p_A\partial_0 A+p_\Phi\partial_0 \Phi-{\cal L}_1\nonumber\\
=
\frac{1}{2}p_A^2+\frac{1}{2}\vec\nabla A\cdot\vec\nabla A
+\frac{1}{2}p_\Phi^2+\frac{1}{2}\vec\nabla\Phi\cdot\vec\nabla\Phi+\frac{1}{2}M^2\Phi^2
\equiv {\cal H}_1\,.
\end{equation}
   Let us consider the generating functional of the Green's functions
of the field $A$ in the canonical path integral representation,
\begin{equation}
Z[J] = \int {\cal D}A {\cal D}p_A {\cal D}\Phi {\cal D}p_\Phi\, e^{i \,\int d^4 x
\,\left(p_A\partial_0 A+p_\Phi\partial_0 \Phi-{\cal H}_1 + J\,A \right)}\,,
\label{GenfuncFree1}
\end{equation}
where ${\cal H}_1$ is given in Eq.~(\ref{hamiltonianh1}).
   In the following, we will repeatedly make use of a Gaussian functional integral
of the form
\begin{equation}
\label{gengauss}
\int {\cal D} p\, e^{i \,\int d^4 x \left(-\frac{1}{2}p^2+fp\right)}={\cal N}
e^{i \,\int d^4 x\, \frac{1}{2}f^2}\,,
\end{equation}
   where ${\cal N}$ is an (irrelevant) multiplicative factor and $f$ a given function
not depending on $p$.
   Applying Eq.~(\ref{gengauss}) to the $p_A$ and $p_\Phi$ (functional) integrations in
Eq.~(\ref{GenfuncFree1}) and omitting, as is common practice, the corresponding
multiplicative factors ${\cal N}$ of Eq.~(\ref{gengauss}), yields
\begin{equation}
Z[J] = \int {\cal D}A {\cal D}\Phi \, e^{i \,\int d^4 x\,
\left[{\cal L}_1(A,\Phi) + J\,A \right]}\,, \label{GenfuncFree2}
\end{equation}
where ${\cal L}_1$ is given in Eq.~(\ref{toylagr1}).
   The (full) propagator of the $A$ field is given by
\begin{equation}
i\Delta_A (p)  =  \frac{i}{p^2+i0^+}\,,  \label{toypropagators}
\end{equation}
and the $A$ field describes a massless non-interacting spinless particle.

\section{Field transformations}

   Our line of arguments relies on the principle that, for a given theory,
the physical content of the theory both at the classical as well as the quantum level
should not depend on the choice of variables for describing the physical degrees of freedom.
   We will make use of the free theory described by the simple Lagrangian (Hamiltonian)
of Eq.~(\ref{toylagr1}) [Eq.~(\ref{hamiltonianh1})] and its canonical path integral quantization
described by Eq.~(\ref{GenfuncFree1}).
   The path integral representation of Eq.~(\ref{GenfuncFree2}) is a deduced quantity in
the sense that it is derived from the canonical result of Eq.~(\ref{GenfuncFree1}).
   The Green's functions obtained from Eq.~(\ref{GenfuncFree2}), in particular the
propagator of Eq.~(\ref{toypropagators}), will be taken as reference quantities.
   We will make use of two types of changes of field variables, namely, transformations
without and with time derivatives of a field.
   Using the reference result of Sec.~\ref{toymodel}, we will be able to point out
that the canonical path integral quantization applied to a Hamiltonian
based on the Ostrogradsky method, at the quantum level, does not describe an equivalent
theory.

\subsection{Field transformation without a time derivative}

   We first consider a change of field variables
involving both spatial derivatives of a new field $\phi$ and the product of
the $A$ field with the $\phi$ field,
\begin{equation}
\Phi(x)= \phi(x) - c\, \Delta \phi(x)+g\,\phi(x) A(x)\,,
\label{toyfieldtransf1}
\end{equation}
where $(x)$ stands for $(t,\vec x)$.
   In Eq.~(\ref{toyfieldtransf1}), the real parameters $c$ and $g$
carry the dimensions of a squared inverse mass and an inverse mass,
respectively, and $\Delta$ denotes the usual Laplace operator.
   The Lagrangian density ${\cal L}_2$ in the new variables is obtained
by substituting $\Phi(x)$ of Eq.~(\ref{toyfieldtransf1}) into the
original Lagrangian ${\cal L}_1$ of Eq.~(\ref{toylagr1}),
\begin{eqnarray}
{\cal L}_2(A,\phi)={\cal L}_1(A,\Phi)\,.
\label{toylagr2}
\end{eqnarray}
   Applying the change of variables, given by
Eq.~(\ref{toyfieldtransf1}), directly to the generating functional
of Eq.~(\ref{GenfuncFree2}), we obtain
\begin{eqnarray}
Z[J] & = & \int {\cal D}A {\cal D}\phi \,
\det\left(\frac{\delta\Phi(y)}{\delta\phi(z)}\right)
 e^{i \,\int d^4 x\, \left[{\cal L}_2(A,\phi) + J\,A \right]}\,,
\label{GenfuncFree3}
\end{eqnarray}
where
\begin{equation}
\label{jacobimatrix}
\left(\frac{\delta\Phi(y)}{\delta\phi(z)}\right)=
\left(
\vphantom{\frac{\delta\Phi(y)}{\delta\phi(z)}}
[1- c\, \Delta_y+g\,A(y)]\delta^4(y-z)\right)
\end{equation}
denotes the Jacobian ``matrix'' of the field transformation.
   Equation (\ref{GenfuncFree3}) is the generalization of the
substitution rule for multiple Riemann integrals to functional integrals.

   The same result as Eq.~(\ref{GenfuncFree3}) for the generating
functional is obtained by first applying the Hamilton
formalism to the Lagrangian ${\cal L}_2$ of Eq.~(\ref{toylagr2})
and by subsequently performing the canonical path integral quantization.
   To that end we define the canonical momenta
\begin{eqnarray}
p_A & = & \frac{\partial {\cal L}_2}{\partial\, \partial_0 A}
= \partial_0 A + g\,\phi\, \partial_0 \Phi \,,
\label{canmompA2}\\
p_\phi  & = &\frac{\partial {\cal L}_2}{\partial\, \partial_0 \phi}
=\left(1 - c\,\Delta+g\,A\right)\partial_0\Phi
\,, \label{canmomphi2}
\end{eqnarray}
where
\begin{equation}
\label{Phidot}
\partial_0\Phi=\left(1-c\,\Delta + g\,A\right)\partial_0\phi+g\,\phi\, \partial_0 A\,.
\end{equation}
   From Eq.~(\ref{canmomphi2}) we obtain
\begin{equation}
\partial_0\Phi=\widehat O p_\phi\equiv \left(1-c\,\Delta+g \,A\right)^{-1} p_\phi\,.
\label{dotPhi2}
\end{equation}
   Substituting Eq.~(\ref{dotPhi2}) into Eq.~(\ref{canmompA2}), we can solve
\begin{equation}
\label{dotAsolve}
\partial_0 A= p_A-g\,\phi\, \widehat O p_\phi\,.
\end{equation}
   Finally, inserting Eqs.~(\ref{dotPhi2}) and (\ref{dotAsolve}) into Eq.~(\ref{Phidot}),
we can solve
\begin{equation}
\partial_0\phi=\left(1-c\,\Delta+g\,A\right)^{-1}
\left\{\left[1+(g\,\phi)^2\right]\widehat O p_\phi-g\,\phi\, p_A\right\}\,.
\label{dotphisolve}
\end{equation}
   Using integration by parts and omitting total divergences, the
Hamiltonian takes the form
\begin{eqnarray}
{\cal H}_2 & = & p_A \partial_0 A+ p_\phi
\partial_0\phi -{\cal L}_2 \nonumber\\
& = & \frac{1}{2} p_A^2+\frac{1}{2}\left[1+(g\,\phi)^2\right]\left(\widehat{O} p_\phi\right)^2
-g\,\phi\, \widehat{O}p_\phi\, p_A\nonumber\\
&&+\frac{1}{2}\vec\nabla A\cdot\vec\nabla A+\frac{1}{2}\vec\nabla\Phi\cdot \vec\nabla\Phi
+\frac{1}{2}M^2\Phi^2\,.
\end{eqnarray}
   In terms of the Hamiltonian ${\cal H}_2$ the generating functional for
Green's functions of the $A$ field is given by \cite{gitman}
\begin{equation}
Z[J] = \int {\cal D}A {\cal D}p_A {\cal D}\phi {\cal D}p_\phi\,
e^{i \int d^4 x\, (p_A\partial_0 A +p_\phi
\partial_0\phi - {\cal H}_2+J\,A)}\,. \label{GenfunctHamilton2}
\end{equation}
   Introducing the new variable $\pi_\phi = \widehat O p_\phi$, we
obtain
\begin{equation}
Z[J]=\int {\cal D} A {\cal D} p_A {\cal D}\phi {\cal
D}\pi_\phi\,\det \left(\frac{\delta p_\phi(y)}{\delta \pi_\phi(z)}\right)
e^{i \int d^4 x \left( p_A\partial_0 A
+\widehat{O}^{-1}\pi_\phi
\partial_0\phi - \tilde{\cal H}_2+J\,A\right)}\,, \label{GenfunctHamilton2new}
\end{equation}
where $\widehat{O}^{-1}\pi_\phi=\left(1-c\,\Delta+g\,A\right)\pi_\phi$.
   The Jacobian matrix is given by
\begin{displaymath}
\left(\frac{\delta p_\phi(y)}{\delta\pi_\phi(z)}\right)=
\left(\vphantom{\frac{\delta p_\phi(y)}{\delta\pi_\phi(z)}}
[1- c\, \Delta_y+g\,A(y)]\delta^4(y-z)\right),
\end{displaymath}
and coincides with the Jacobian matrix of Eq.~(\ref{jacobimatrix}).
   Finally, the Hamiltonian $\tilde {\cal H}_2$ reads
\begin{eqnarray}
\tilde {\cal H}_2 & = & \frac{1}{2}\left(p_A - g\,\phi\,\pi_\phi
\right)^2+\frac{1}{2}\pi_\phi^2
+\frac{1}{2}\vec\nabla A\cdot\vec\nabla A+\frac{1}{2}\vec\nabla\Phi\cdot \vec\nabla\Phi
+\frac{1}{2}M^2\Phi^2\,.
\label{hamiltonian2tilde}
\end{eqnarray}
   By means of partial integration in the exponent of Eq.~(\ref{GenfunctHamilton2new}),
the expression $\widehat{O}^{-1}\pi_\phi\partial_0\phi$ is replaced
by $\pi_\phi\widehat{O}^{-1}\partial_0\phi$.
   Performing subsequently the $p_A$ and $\pi_\phi$ integrations using Eq.~(\ref{gengauss}),
we obtain
\begin{equation}
Z[J]= \int {\cal D}A {\cal D}\phi \,\det \Big([1- c \,\Delta_y+g\,
A(y)] \,\delta^4(y-z)\Big)\, e^{i \int d^4 x\, \left[ {\cal L}_2(A,\phi)+J\,A\right]}\,,
\label{GenfunctHamilton2final}
\end{equation}
which is identical with Eq.~(\ref{GenfuncFree3}).
   To summarize this section, given the change of variables of Eq.~(\ref{toyfieldtransf1})
(without a time derivative), the substitution in the functional integral
of Eq.~(\ref{GenfuncFree2}) yields the same result as the application of the canonical
path integral quantization starting from the Hamiltonian ${\cal H}_2$ derived from
the Lagrangian ${\cal L}_2$.

\subsection{Field transformation with time derivatives}
   We now go one step further and consider the following change of field variables
involving time derivatives,
\begin{equation}
\Phi(x)=\chi(x)+c\, \Box \chi(x)+g\,\chi(x)\, A(x)\,,
\label{toyfieldtransf}
\end{equation}
where $\Box=\partial_0^2-\Delta$ denotes the d'Alembert operator.
   The Lagrangian in the new variables is obtained from
\begin{eqnarray}
{\cal L}_3(A,\chi) & = & {\cal L}_1(A,\Phi)
\,.\label{toylagr3}
\end{eqnarray}
   Because of the d'Alembertian in the field transformation, the Lagrangian
${\cal L}_3(A,\chi)$ contains time derivatives of the field $\chi$ up to
and including third order.
   Performing the change of variables in the generating functional of
Eq.~(\ref{GenfuncFree2}) results in
\begin{equation}
Z[J]=\int {\cal D}A {\cal D}\chi \,\det \left(\frac{\delta\Phi(y)}{\delta\chi(z)}\right)
 e^{i \,\int d^4 x\, \left[{\cal L}_3(A,\chi) + J\,A \right]}\,, \label{GenfuncFree4}
\end{equation}
with the Jacobian matrix
$$
\left(\frac{\delta\Phi(y)}{\delta\chi(z)}\right)=
\left(\vphantom{\frac{\delta\Phi(y)}{\delta\chi(z)}}
[1+ c \,\Box_y+g\,A(y)]\delta^4(y-z)\right).
$$
   We express the determinant of the Jacobian matrix in terms of a
functional integral over ghost fields $g_1$ and $g_2$ (scalar Grassmann variables),
\begin{equation}
\det\,\left(\vphantom{\frac{\delta\Phi(y)}{\delta\chi(z)}}
[1+ c \,\Box_y+g\,A(y)]\delta^4(y-z)\right)
=\int {\cal D}g_1{\cal D}g_2\, e^{i\,\int d^4 x\,g_2(1+c\,\Box+g\,A)g_1}\,.
\end{equation}
   In this representation the
generating functional takes the following form,
\begin{equation}
Z[J] = \int {\cal D}A {\cal D}\chi {\cal D}g_1{\cal D}g_2\,
e^{i \,\int d^4 x\, \left[{\cal L}_3(A,\chi) +g_2\,\left(1+ c\,
\Box+g\, A\right)g_1+ J\,A \right]}\,.
\label{GenfunctLagrangeFinal}
\end{equation}
   Using the example of the full propagator of the $A$ field, we will illustrate
that Eq.~(\ref{GenfunctLagrangeFinal}) gives rise to the same Green's functions
including quantum corrections.

\subsection{Full propagator of the $A$ field at the one-loop level}
   We will investigate the propagator of the $A$ field resulting from the
perturbative expansion of Eq.~(\ref{GenfunctLagrangeFinal}) at
the one-loop level.
   We will explicitly see that the quantum corrections obtained from Eq.~(\ref{GenfunctLagrangeFinal})
do not modify the position of the pole, i.e., the $A$ field remains massless.
   The dressed propagator of the $A$ field is of the form
\begin{equation}
i \Delta_A(p)= \frac{i}{p^2-\Sigma (p^2)}\,, \label{dressedAprop}
\end{equation}
where $-i\Sigma$ denotes the proper self-energy insertions of the $A$ field, i.e.,
the sum of one-particle-irreducible diagrams contributing to the two-point
function.

\begin{figure}
\epsfig{file=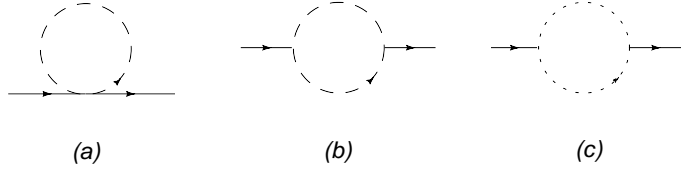, width=9truecm}
\caption[]{\label{mnloops:fig} One-loop contributions to the
self energy of the $A$ field.
The solid, dashed, and dotted lines correspond to $A$, $\chi$,  and the ghost fields,
respectively.}
\end{figure}

   At the one-loop level, three diagrams contribute to the self energy
(see Fig.~\ref{mnloops:fig}).
   The corresponding Feynman rules are summarized in the Appendix.
   Using dimensional regularization, we obtain at $p^2=0$,
\begin{equation}
\Sigma^{(a)}(0)= -ig^2 \int \frac{d^n
k}{(2 \pi)^n}\frac{1}{(1-c\,k^2)^2}
=-\frac{1}{2}\Sigma^{(b)}(0)=\Sigma^{(c)}(0)\,.
\label{sediagrams}
\end{equation}
   These contributions cancel each other and, as a result, the quantum
corrections do not give rise to a mass of the $A$ field.
   This was expected, as the field transformation cannot change the physical
content of a theory.

\section{Ostrogradsky's Hamilton formalism}

   We now apply the Hamilton formalism and the canonical
quantization to the Lagrangian of Eq.~(\ref{toylagr3}) and derive
the generating functional.
   As the Lagrangian ${\cal L}_3$ contains time derivatives of higher
orders, we use the Ostrogradsky formalism.
   While exactly the same results are obtained by following the
procedure of Ref.~\cite{gitman}, here we apply the method
of Ref.~\cite{Govaerts:1994zh} based on an auxiliary Lagrangian.
    We first define new independent fields
\begin{eqnarray}
\psi & = & \partial_0 \chi\,,\nonumber\\
\zeta & = & \partial_0 \psi\,.\label{defPsi2}
\end{eqnarray}
   Introducing the Lagrange multipliers $\lambda_1$ and $\lambda_2$
in order to enforce the relations of Eq.~(\ref{defPsi2}), we obtain
the auxiliary Lagrangian
\begin{eqnarray}
{\cal L}_{\rm aux} & = & \frac{1}{2}\,\partial_\mu A\partial^\mu A
+\frac{1}{2}\,\left[\psi+c\,(\partial_0\zeta-\Delta\psi)+g\,\partial_0 A\,\chi
+g\,A\,\psi\right]^2\nonumber\\
&&-\frac{1}{2}\,\vec\nabla\left[\chi+c\,(\zeta-\Delta\chi)+g\,A\,\chi\right]\cdot
\vec\nabla\left[\chi+c\,(\zeta-\Delta\chi)+g\,A\,\chi\right]\nonumber\\
&&-\frac{1}{2}\,M^2\left[\chi+c\,(\zeta-\Delta\chi)+g\,A\,\chi\right]^2
+\lambda_1(\psi-\partial_0\chi)+\lambda_2(\zeta-\partial_0\psi)\,.
\end{eqnarray}
    The momenta canonically conjugated to the degrees of freedom
$A$, $\chi$, $\psi$, $\zeta$, $\lambda_1$, and $\lambda_2$ are defined
as
\begin{eqnarray}
p_A&=&\frac{\partial{\cal L}_{\rm aux}}{\partial\, \partial_0 A}=
\partial_0 A \left(1+ g^2\,  \chi ^2\right) +g\, \chi\,\left[
\psi+c\,(\partial_0\zeta
-\Delta \psi)+g\,A\, \psi \right] \,,\nonumber\\
p_\chi&=&\frac{\partial{\cal L}_{\rm aux}}{\partial\, \partial_0 \chi}=-\lambda_1\,,\nonumber\\
p_\psi&=&\frac{\partial{\cal L}_{\rm aux}}{\partial\, \partial_0 \psi}=-\lambda_2\,,\nonumber\\
p_\zeta&=&\frac{\partial{\cal L}_{\rm aux}}{\partial\, \partial_0 \zeta}
=c\, \left[\psi+ c\, (\partial_0\zeta -\Delta \psi)+g\,
\partial_0 A\,\chi +g\, A\, \psi\right] \,,\nonumber\\
p_{\lambda_1}&=&\frac{\partial{\cal L}_{\rm aux}}{\partial\, \partial_0 \lambda_1}=0\,,\nonumber\\
\label{canmomenta}
p_{\lambda_2}&=&\frac{\partial{\cal L}_{\rm aux}}{\partial\, \partial_0 \lambda_2}=0\,.
\end{eqnarray}
   For $c\neq 0$, the two equations for $p_A$ and $p_\zeta$ can be inverted to
solve $\partial_0 A$ and $\partial_0\zeta$, respectively,
\begin{eqnarray*}
\partial_0 A & = & p_A - \frac{1}{c}\,g\,\chi\,p_\zeta \,,\\
\partial_0 \zeta & = &\frac{1}{c^2}\left\{
\left[1+(g\,\chi)^2\right]p_\zeta-c\,g\,\chi\,p_A-c\,(1-c\,\Delta +g\,A)\psi\right\}\,.
\end{eqnarray*}
The remaining velocities cannot be solved from Eqs.~(\ref{canmomenta}), i.e., the corresponding
momenta need to satisfy the primary constraints
\begin{eqnarray}
\Phi_1 & = & p_{\lambda_1}\approx 0\,,\nonumber\\
\Phi_2 & = & p_{\lambda_2}\approx 0\,,\nonumber\\
\Phi_3 & = & p_\chi+\lambda_1\approx 0\,,\nonumber\\
\Phi_4 & = & p_\psi+\lambda_2\approx 0\,. \label{fourconstraints}
\end{eqnarray}
   Here, $\Phi_i\approx 0$ denotes a weak equation in Dirac's
sense, namely that one must not use one of these constraints before
working out a Poisson bracket \cite{Dirac}.
   The so-called total or generalized Hamiltonian ${\cal H}^{(1)}$ has the form
\begin{equation}
\label{hamden}
{\cal H}^{(1)}=\sum_{i=1}^4 \Phi_i z_i+{\cal H}\,,
\end{equation}
where
\begin{eqnarray}
\label{parthamiltonian}
{\cal H} & = & \frac{1}{2}\left[p_A-\frac{g\,\chi}{c}p_\zeta\right]^2
+\frac{1}{2}\,\frac{p_\zeta^2}{c^2}
-\frac{1}{c}\,p_\zeta(1-c\,\Delta+g\,A)\psi+\frac{1}{2}\,\vec\nabla A\cdot\vec\nabla A\nonumber\\
&&+\frac{1}{2}\,\vec\nabla\left[\chi+c\,(\zeta-\Delta\chi)+g\,A\,\chi\right]\cdot
\vec\nabla\left[\chi+c\,(\zeta-\Delta\chi)+g\,A\,\chi\right]\nonumber\\
&&+\frac{1}{2}\,M^2\left[\chi+c\,(\zeta-\Delta\chi)+g\,A\,\chi\right]^2
-\lambda_1\psi-\lambda_2\zeta\,.
\end{eqnarray}
   In Eq.\ (\ref{hamden}), the $z_i$ are arbitrary functions which have to
be determined.
   The constraints of Eq.~(\ref{fourconstraints}) have to be conserved in
time.
   Therefore, we demand that the Poisson brackets of $\Phi_i$ with
$H^{(1)}=\int d^3 x \,{\cal H}^{(1)}$ vanish.
   An explicit evaluation of the Poisson brackets yields
\begin{eqnarray*}
\left\{\Phi_1,H^{(1)}\right\}\approx0\,\Rightarrow\, z_3&=&\psi\,,\\
\left\{\Phi_2,H^{(1)}\right\}\approx0\,\Rightarrow\, z_4&=&\zeta\,,\\
\left\{\Phi_3,H^{(1)}\right\}\approx0\,\Rightarrow\, z_1
&=&-\frac{g}{c}\,p_\zeta\left(p_A-\frac{g\,\chi}{c}\,p_\zeta\right)\\
&&-(1-c\,\Delta+g\,A)(\Delta-M^2)(\chi+c\,\zeta-c\,\Delta\,\chi+g\,\chi\,A)\,,\\
\left\{\Phi_4,H^{(1)}\right\}\approx0\,\Rightarrow\,
z_2&=&-\frac{1}{c}\,(1-c\,\Delta+g\,A)p_\zeta-\lambda_1\,.
\end{eqnarray*}
   According to Ref.~\cite{gitman}, the generating functional for the Green's
functions of the $A$ field can be written as a path
integral over canonical coordinates and momenta,
\begin{eqnarray}
Z[J] & = & \int {\cal D}A {\cal D}p_A {\cal D}\chi {\cal D}p_\chi
{\cal D}\lambda_1 {\cal D}p_{\lambda_1}
{\cal D}\lambda_2 {\cal D}p_{\lambda_2}
{\cal D}\psi {\cal D}p_\psi {\cal D}\zeta {\cal D}p_\zeta\,\nonumber\\
&&\times
\delta\left[\Phi_1\right]\delta\left[\Phi_2\right]\delta\left[\Phi_3\right]\delta\left[\Phi_4\right]
\left[\det\left( \{\Phi,\Phi\}\right)\right]^\frac{1}{2}\, e^{i
{\cal S}[J]}\,, \label{GenfunctHamilton}
\end{eqnarray}
where
\begin{equation}
{\cal S}[J]=
\int d^4 x \left(p_A\partial_0 A+p_\chi\partial_0\chi
+p_{\lambda_1}\partial_0\lambda_1+p_{\lambda_2}\partial_0\lambda_2
+p_\psi \partial_0 \psi +p_\zeta \partial_0\zeta - {\cal H}+J\,A\right)\,. \nonumber \label{action}
\end{equation}
   The delta functionals in Eq.~(\ref{GenfunctHamilton}) take care of the
constraints of Eq.~(\ref{fourconstraints}).
   The entries of the $4\times 4$ matrix $\left(\{\Phi,\Phi\}\right)$ are given
by the Poisson brackets $\{\Phi_i,\Phi_j\}$ and, in the present case, the
determinant reduces to a constant which will be omitted in the following.

   In Eq.~(\ref{GenfunctHamilton}), because of the delta functionals, the integrations
over $\lambda_i$ and $p_{\lambda_i}$ are straightforward and give rise to the intermediate
result
\begin{eqnarray}
Z[J] & = & \int {\cal D}A {\cal D}p_A {\cal D}\chi {\cal D}p_\chi
{\cal D}\psi {\cal D}p_\psi {\cal D}\zeta {\cal D}p_\zeta\,\nonumber\\
&&\times
e^{i \,\int d^4 x\left(p_A\partial_0A+p_\chi\partial_0\chi+p_\psi\partial_0\psi
+p_\zeta\partial_0\zeta
-\frac{1}{2}\left[p_A-\frac{g\,\chi}{c}p_\zeta\right]^2-\dots
-\frac{1}{2}M^2\left[\chi+c\,(\zeta-\Delta\chi)+g\,A\,\chi\right]^2
-p_\chi\psi-p_\psi\zeta+JA
\right)}\,.\nonumber\\
\label{Zintermediate}
\end{eqnarray}
   The momenta $p_\chi$ and $p_\psi$ appear linearly in the exponent of
Eq.~(\ref{Zintermediate}).
   Therefore, the integrations over $p_\chi$ and $p_\psi$ give rise to
the delta functionals $\delta[\partial_0\chi-\psi]$, and
$\delta[\partial_0\psi-\zeta]$, respectively.
    The $\zeta$ integration then results in the replacement $(\zeta,\partial_0\zeta)\to
(\partial_0\psi,\partial_0^2\psi)$, and the $\psi$ integration in
$(\psi,\partial_0\psi,\partial_0^2\psi)\to(\partial_0\chi,\partial_0^2\chi,\partial_0^3\chi)$.
   Finally, using Eq.~(\ref{gengauss}) to perform the $p_A$ and $p_\zeta$ integrations,
and inserting Eq.~(\ref{toyfieldtransf}),
we obtain the following expression,
\begin{eqnarray}
Z[J] &=&\int {\cal D}A {\cal D}\chi\, e^{i \,\int d^4 x\,
\left[\frac{1}{2}(\partial_0A)^2-\frac{1}{2}\vec\nabla A\cdot\vec\nabla A
+\frac{1}{2}(\partial_0\Phi)^2-\frac{1}{2}\vec\nabla\Phi\cdot\vec\nabla\Phi
-\frac{1}{2}M^2\Phi^2+J\,A\right]}\nonumber\\
&=&
\int {\cal D}A {\cal D}\chi\, e^{i \,\int d^4 x\, \left[{\cal
L}_3(A,\chi) + J\,A \right]}\,. \label{GenfunctLagrange}
\end{eqnarray}
   A comparison of Eq.~(\ref{GenfunctLagrange}) with
Eq.~(\ref{GenfunctLagrangeFinal}) shows that the two results
differ in the ghost part of the effective Lagrangian.
   Considering the dressed propagator of
the $A$ field generated by Eq.~(\ref{GenfunctLagrange}) we see
that only the diagrams (a) and (b) in Fig.~\ref{mnloops:fig}
contribute and hence their contributions at $p^2=0$ do not cancel.
   As a result the field $A$ gains a non-vanishing mass due to the quantum
corrections.
   This evidently contradicts the original physical
content of the considered toy model.

\section{Conclusions}

   In this work we discussed the quantization of field theories
containing higher-order time derivatives of the fields in
the Lagrangian.
   We started from a model describing one massless and one massive free
spinless particle.
   The corresponding path-integral representation of the
generating functional for the Green's functions
of the massless field served as the reference point.
   We performed a change of field variables involving (time)
derivatives resulting in a new Lagrangian containing higher-order
time derivatives.
   To this Lagrangian we applied Ostrogradsky's Hamilton formalism
for theories with higher derivatives and subsequently quantized
the obtained theory using canonical quantization.
   The resulting generating functional for the Green's
functions of the massless field differs from the one obtained by
the change of variables in the reference generating functional.
   As a specific consequence, we showed that Ostrogradsky's formalism
gives rise to a non-vanishing mass contribution for the massless
particle due to the quantum corrections.
   These findings may be understood as follows.
   Ostrogradsky's formalism is equivalent to the introduction of new
non-physical degrees of freedom.
   At the classical level, for the field variables of the original
Lagrangian, Ostrogradsky's formalism leads to equations of motion
which are equivalent to the ones of the Lagrange formalism.
   On the other hand, the non-physical degrees of freedom result in
non-trivial contributions at the level of quantum corrections.
   We conclude that Ostrogradsky's
Hamilton formalism may lead to wrong results and therefore, in
general, cannot be considered as a satisfactory basis for the
quantization of systems described with Lagrangians involving
higher derivatives.

\acknowledgments

This work was supported by the Deutsche Forschungsgemeinschaft
(SFB 443).

\section{Appendix}
   Up to a total divergence, the Lagrangian of Eq.~(\ref{toylagr3}) can be written as
\begin{eqnarray}
\label{L3explicit}
{\cal L}_3(A,\chi)&=&\frac{1}{2}\partial_\mu A\partial^\mu A
-\frac{1}{2}(\chi+c\,\Box\chi)(\Box+M^2)(\chi+c\,\Box\chi)
\nonumber\\
&&-g\,A\,\chi (\Box+M^2)(\chi+c\,\Box\chi)
+\frac{1}{2}\,g^2\left[A^2(\partial_\mu\chi\partial^\mu\chi-M^2\chi^2)-\chi^2 A\,\Box A\right]\,.
\end{eqnarray}
   In combination with the ghost contribution, Eq.~(\ref{L3explicit}) results in
the following Feynman rules:
\begin{enumerate}
\item Internal line of an $A$ field with momentum $k$:
\begin{displaymath}
\frac{i}{k^2+i0^+}\,.
\end{displaymath}
\item Internal line of a $\chi$ field with momentum $k$:
\begin{displaymath}
\frac{i}{(k^2-M^2+i0^+)(1-c\,k^2)^2}\,.
\end{displaymath}
\item Internal line from ghost fields with momentum $k$:
\begin{displaymath}
\frac{i}{1-c\,k^2}\,.
\end{displaymath}
\item Vertex $\chi(p_i)\to \chi(p_f)+A$:
$ig[(1-c\,p_i^2)(p_i^2-M^2)+(1-c\,p_f^2)(p_f^2-M^2)].$
\item Vertex $\chi(p_i)+A(k_i)\to\chi(p_f)+A(k_f)$:
$ig^2[2(p_i\cdot p_f-M^2)+k_i^2+k_f^2].$
\item Vertex $g_1\to g_2+A$:
$ig$.
\item Because of the Grassmann nature of the ghost fields, a ghost loop produces an overall minus sign.
\end{enumerate}


\end{document}